\documentclass{ceab}   

\usepackage{epsfig}     
\usepackage{graphicx}   
\usepackage{amssymb}
\usepackage{rotating}

\usepackage{color}

\usepackage{url}

\usepackage{ceabbib}     
\usepackage[T1]{fontenc}

\begin{document}

\def\tit{Radio signatures of SEP events}
\def\aut{Miteva et al.}
\def\str{??--??}

\title{Radio signatures of solar energetic particles during the 23$^{\rm rd}$ solar cycle}

\author{R. MITEVA$^{1}$, K.-L. KLEIN$^1$, S.~W. SAMWEL$^2$, A. NINDOS$^3$, \\ A. KOULOUMVAKOS$^{3,4}$ and H. REID$^{1,5}$\\
\vspace{2mm}\\
\it $^1$ LESIA-Observatoire de Paris, 5 place Jules Janssen, 92195 Meudon, \\
\it CNRS, UPMC, Univ. Paris-Diderot, France\\
\it $^2$ National Research Institute of Astronomy and Geophysics (NRIAG), \\
\it Helwan, Cairo, Egypt\\
\it $^3$ Section of Astrogeophysics, Physics Department, University of Ioannina, \\
\it Ioannina GR-45110, Greece \\
\it $^4$ Department of Astrophysics, Astronomy and Mechanics, Faculty of Physics, \\
\it University of Athens, Athens, Greece \\
\it $^5$ SUPA School of Physics and Astronomy, University of Glasgow, \\
\it Glasgow G12 8QQ, UK \\
}

\maketitle

\begin{abstract}
We present the association rates between solar energetic particles (SEPs) and the radio emission signatures in
the corona and IP space during the entire solar cycle 23. We selected SEPs associated with X and M-class flares from the visible solar hemisphere. All SEP events are also accompanied by coronal mass ejections. Here, we focus on the correlation between the SEP events and the appearance of radio type II, III and IV bursts on dynamic spectra. For this we used the available radio data from ground-based stations and the Wind/WAVES spacecraft. The associations are presented separately for SEP events accompanying activity in the eastern and western solar hemisphere. We find the highest association rate of SEP events to be with type III bursts, followed by types II and IV. Whereas for types III and IV no longitudinal dependence is noticed, these is a tendency for a higher SEP-association rate with type II bursts in the eastern hemisphere. A comparison with reports from previous studies is briefly discussed.

\end{abstract}


\section{Introduction}

Solar energetic particle (SEP) events are transient flux enhancements of electrons, protons and ions due to acceleration processes in the solar corona and interplanetary (IP) space. The high energy particles can pose a serious risk for the near Earth and ground-based technological devices, may disturb communications and be a health hazard \citep{2006SSRv..124..303M}. This is why energetic particles are of major space weather interest. The accelerator of these particles, however, is still a subject of debate. Presently, the reconnection processes during flares \citep{2002JGRA..107.1315C} and the shock-acceleration at coronal mass ejections (CMEs), \cite{1999SSRv...90..413R}, are recognized as efficient particle accelerators. An important issue is whether both acceleration processes act simultaneously with comparable efficiency or one of them dominates the particle energization process. In the current two-class SEP picture, the flares are thought to dominate the {\it impulsive} events and the shock waves dominate the large {\it gradual} events. Observations, however, are not able to relate unambiguously the SEP parameters measured in situ to the parent solar activity (e.g., flares vs. shocks). The large uncertainties usually present when doing different correlation analyses may arise due to the poorly known particle transport in the turbulent IP magnetic field, the magnetic connection between the acceleration site and the Earth, and that SEPs are often just measured at a single point in space.

Radio observations can provide additional information and constraints for identifying the particle accelerator in the corona and IP space. When nonthermal electrons propagate through the solar corona they may emit radio waves, if certain conditions apply \citep{2008SoPh..253....3N}. The different types of radio bursts in the dynamic radio spectrum (see e.g., a review by \citet{2008A&ARv..16....1P}), are usually interpreted as signatures from electrons accelerated in the vicinity of a shock wave (type IIs), as electron beams (type IIIs) or as electrons confined in closed loop structure during the early evolution of a CME (type IVs). Type III bursts usually extend to low frequencies (reaching as far as the Earth orbit, $\sim$20 kHz) which is the main indicator that accelerated electrons (and by inference also protons) can efficiently escape from the solar corona \citep{2002JGRA..107.1315C}. Interplanetary counterparts of type II are also observed, but to lesser extent, whereas type IVs can be seen only to few MHz, but not to lower frequencies. Here we focus on the comparison of the particle (mainly proton) intensities measured in situ with the electromagnetic emission of electrons in the corona and IP space over the entire solar cycle 23.

Numerous previous studies reported different association rates of SEPs and emission of radio bursts of type II \citep{2004ApJ...605..902C,2008AnGeo..26.3033G}, III \citep{2002JGRA..107.1315C} and IV \citep{1982ApJ...261..710K}. The aim here is to identify the different types of radio emission from the dynamic spectra, to complement our results with reports from the different radio observatories and to compare them with previous works on the topic.

\section{Particle data}

In the present work we selected all proton events with energy above 25 MeV as identified by \cite{2010JGRA..11508101C} that are associated with both strong flares (X and M-class) at eastern and western heliolongitudes and CMEs. Here we will differentiate the SEP events into eastern/western only and not by other SEP classifications schemes \citep{1999SSRv...90..413R,2009CEAB...33..253C}, since any classification may leave out many `mixed' cases in abundances, charge states, associated phenomena, etc. However, in order to facilitate comparison with previous work, we note if the SEP event was described as `gradual' or `impulsive' in \cite{2004ApJ...610..510R} and \cite{2009ApJ...690..598C} by the superscripts `g' and `i', respectively.

In order to improve the statistics for the radio analysis (see next Section) we included SEP events that have high background level due to a previous event\footnote{Several events with high background are not present in the list of \cite{2010JGRA..11508101C} but are adopted from \cite{2011SoPh..269..309K}, noted with superscript `n'.}, that are observed during SOHO data gap, those for which no value for the peak intensity was given due to instrument saturation (with superscript `s') and those with parent activity at the limb (with superscript `l' we denote events at the solar limb and with `c', events close to the disc center, between $\pm 10$ degrees in heliolongitude). That lead finally to 175 particle events during solar cycle 23, of which 49 had sources in the eastern hemisphere, 124 in the western and two had uncertain source locations.

\section{Radio spectrograph data}

The radio emission observed on ground and in space is usually presented in a frequency vs. time plot, where the strength of the radio emission is color-coded. This is known as a dynamic radio spectrum. Several features were recognized on such spectrum plots, e.g., fast drifting emission stripes extending from high to low frequencies (type IIIs); slowly drifting lanes of emission (type IIs) and a broad band stationary or/and drifting emission of type IV. Each of these radio emission types is a result of a unstable electron population (produced by a different process) generating Langmuir waves that convert into electromagnetic radiation via wave-wave processes. Namely, type IIs are usually assumed to be the shock signatures in the corona/IP space \citep{1985srph.book..333N}, type IIIs are electron beams propagating through the corona \citep{1985srph.book..289S}, and the type IVs are the signatures from trapped electrons in coronal loops \citep{1985srph.book..361S,1986SoPh..104...19P}. Here, we will use this standard interpretation for the radio burst emission in order to identify the probable particle accelerator. As a preparatory work for the analysis we collected all available radio spectral data (summarized in Table~\ref{T-obslist}) and the associated GOES soft X-ray (SXR) emission for each SEP event. The results of the associated radio bursts to each particle event are summarized in Tables~\ref{T-Western_events1} and \ref{T-Eastern_events}. There, we start with the SEP event date followed by the onset (in UT) of the SXR emission associated with each event as provided by GOES satellite (1$-$8 \AA$\,$ channel).

\begin{table}[t!]
\caption{List of radio data sources}
\label{T-obslist}
\tiny
\begin{center}
\begin{tabular}{lll}
\hline
Observatory & Frequency range & Data link \\
\hline
Phoenix-2           & 100$-$4000 MHz & \url{http://soleil.i4ds.ch/solarradio/} \\
HiRAS               & 25$-$2500 MHz  & \url{http://sunbase.nict.go.jp/solar/denpa/index.html} \\
Culgoora            & 18$-$1800 MHz  & \url{http://www.ips.gov.au/World_Data_Centre/1/9}\\
Ondrejov            & 800$-$4500 MHz & \url{http://www.asu.cas.cz/~radio/info.htm} \\
Potsdam             & 40$-$800 MHz   & \url{http://ooo.aip.de/groups/osra/data/montab/} \\
Artemis             & 20$-$650 MHz   & \url{http://web.cc.uoa.gr/~artemis/Artemis4_list.html} \\
Izmiran             & 25$-$270 MHz   & \url{http://www.izmiran.ru/stp/lars/} \\
Learmonth           & 18$-$180 MHz   & \url{http://www.ips.gov.au/World_Data_Centre/1/9}\\
RSTN                & 18$-$180 MHz   & \url{ftp://ftp.ngdc.noaa.gov/STP/SOLAR_DATA/SOLAR_RADIO} \\
DAM                 & 20$-$75 MHz    & \url{http://bass2000.obspm.fr/home.php} \\
Green Banks         & 18$-$70, 170$-$1070 MHz& \url{http://www.astro.umd.edu/~white/gb/index.shtml}\\
Wind/WAVES          & 0.02$-$14 MHz  & \url{http://www-lep.gsfc.nasa.gov/waves/data_products.html} \\
\hline
\end{tabular}
\end{center}
\end{table}

The so-identified radio emissions of type II, III and IV are organized in several frequency (wavelength) ranges in Tables~\ref{T-Western_events1} and \ref{T-Eastern_events}. Namely, the decimeter (dm) range is subdivided into high (0.8$-$3 GHz) and lower (0.3$-$0.8 GHz) frequency parts. Similarly, we divided the metric (m) range into 100$-$300 and 30$-$100 MHz subbands. The IP space (dekameter/hecto\-meter, DH, and kilometer wavelengths) is represented by one column. The radio bursts were ordered by their type and not by their temporal appearance on the radio spectral plot.

Since we primarily used quicklook radio spectral data where image quality may be low, we also collected all available radio observatory reports for each event. In case radio emission of a given type was reported but could not be identified by us (due to low resolution of the actual image or because no radio spectrum plot was found), we give the result in squared brackets in Tables~\ref{T-Western_events1} and \ref{T-Eastern_events}. Any uncertain radio burst identification is indicated by a question mark following the roman number of the corresponding radio burst type. Weak emission signatures are denoted with superscript `w', delayed emission with `d' and low (high)-frequency emission onset with `LFo' (`HFo'). Unclassified emission is given with `UNCLF', fine structures with `FS' and fundamental-harmonic emission with `FH'. Continuum (`CONT') and decimeter (`DCIM') emission in the 0.3$-$3 GHz range is considered as type IV-like emission in the analysis. The complete particle event list together with the associated radio bursts is given in Tables~\ref{T-Western_events1} and \ref{T-Eastern_events} (for western and eastern events, respectively).

\begin{table}[ht!]
\caption[]{Solar energetic particle events with origin at western heliolongitudes: visual identification and [observatory reports] of type II, III and IV radio bursts.}
\label{T-Western_events1}
\tiny
\vspace{0.1cm}
\begin{tabular}{lllllll}
\hline
Event            & SXR   & \multicolumn{2}{c}{dm-$\lambda$}& \multicolumn{2}{c}{m-$\lambda$} & DH-$\lambda$  \\
yymmdd           & onset &3$-$0.8 GHz    & 0.8$-$0.3 GHz   & 300$-$100 MHz   & 100$-$30 MHz  & 30$-$0.02 MHz      \\
\hline
970521           & 20:08 & III?      & III?, IV        & III, IV         & II$^{\rm FH}$, III, IV  & II$^{\rm FH}$, III    \\
971103           & 10:18 & DCIM      & III, IV$^{\rm FS}$& II$^{\rm FH}$, III, IV  & II$^{\rm FH}$, III, IV  & II$^{\rm FH}$, III    \\
971104           & 05:52 & III       & III             & II$^{\rm FH}$, III, IV  & II$^{\rm FH}$, III, IV & [II], III           \\
971106$^{\rm g}$  & 11:49 & CONT      & III, IV         & II, III, IV     & II?, III, IV  & II?, III        \\
980502           & 13:31 & DCIM      & III, IV         & II?, III, IV    & II, III, IV   & II, III, IV     \\
980506           & 07:58 & DCIM      & III, IV         & II$^{\rm FH}$, III, IV   & II, III       & II, III         \\
980930$^{\rm g}$  & 13:08 & DCIM      & CONT            & [II], III       & [II], III     & [II], III      \\
981105$^{\rm g}$  & 19:00 & no data   & no data         & III, IV         & [II], IV      & [II], III           \\
981122           & 06:30 & UNCLF     & III             & [II], III       & [II], III     & III           \\
981122$^{\rm n}$  & 16:30 & no data   & no data         &  no data        & no data       & III            \\
981217           & 07:40 & [DCIM]    & III             & II, III         & II, III       & III           \\
990604$^{\rm g}$  & 06:52 & DCIM      & DCIM, [III]     & CONT            & II$^{\rm FH}$, III & II$^{\rm FH}$, III$^{\rm w}$, [IV]   \\
990627           & 08:34 & [DCIM]    & [DCIM, III]     & III             & II, III       & III, [IV]       \\
990828           & 17:52 & no data   & no data         & no data         & [II, III]     & [II], III           \\
991228           & 00:39 & CONT      & III, IV?        & II$^{\rm FH}$, III & II$^{\rm FH}$, III  & III \\
000212           & 03:51 & [DCIM]    & IV              & IV              & II, III, IV   & II?, III       \\
000302           & 08:20 & DCIM, [III] & CONT?         & II$^{\rm FH}$, III  & II$^{\rm FH}$, III, IV  & II$^{\rm FH}$, III \\
000303           & 02:08 & III       & III             & II$^{\rm FH}$, III, IV  & II$^{\rm FH}$, III, IV  & III           \\
000322           & 18:34 & no data   & no data         & no data         & [II, IV]      & III$^{\rm d}$ \\
000324           & 07:41 &  DCIM     & CONT            & II, III         & II, III       & [II], III             \\
000404$^{\rm g}$  & 15:12 & DCIM      & III, IV         & II, III, IV     & II, III       & [II], III           \\
000501$^{\rm i}$  & 10:16 & DCIM      & no data         & III             & III           & III$^{\rm w}$    \\
000523$^{\rm i}$  & 20:48 & no data   & no data         & III             & III           & III                      \\
000610           & 16:40 & [DCIM]    & no data         &  no data        & [II, III]     & [II], III                 \\
000615           & 19:38 & no data   & no data         &  no data        & [II]          & [II$^{\rm FH}$], III           \\
000617           & 02:25 & IV?$^{\rm d}$ & III$^{\rm d}$ & III$^{\rm d}$    & III           & [II], III$^{\rm LFo}$, [IV]\\
000618           & 01:52 & CONT?     & CONT?           & II$^{\rm HFo}$, III & II, III    & III                      \\
000623           & 14:18 & DCIM      & [DCIM]          & [DCIM]          & II, IV        & II, III                  \\
000625           & 07:17 & no data   & [DCIM]          & II?$^{\rm w}$ , IV & II, [IV]     & [II$^{\rm w}$], III$^{\rm w}$        \\
000714$^{\rm c,g}$& 10:03 & DCIM      & no data         & II?, III, IV?   & III           & [II], III             \\
000722           & 11:17 & DCIM      & [IV]            & II$^{\rm FH}$, III, [IV]  & II?, III, IV   & [II], III           \\
000812$^{\rm i}$  & 09:45 & DCIM      & [III]           & [III]           & II?, III?     & III$^{\rm w}$        \\
000909           & 08:28 & DCIM, II? & II?, [III], IV? & II$^{\rm FH}$, III   & II$^{\rm FH}$    & III$^{\rm w}$        \\
000912$^{\rm c}$  & 11:31 & no data   & no data         & II$^{\rm FH}$, III   & II, III$^{\rm d}$ & II, III      \\
000919           & 08:06 & DCIM      & [II$^{\rm HFo}$], III, IV?        & II, III, IV     & II, III, IV   & II?, III       \\
001108$^{\rm g,s}$& 22:42 & [IV]      & [IV]            &  [III, IV]      & III, [IV]     & II, III              \\
001124$^{\rm c,g}$& 04:55 & III       & III             & II, III         & II, III       & II, III       \\
001124$^{\rm c}$  & 14:51 & [DCIM, III] & [DCIM, III]   & II, III         & II, III       & II, III       \\
010128$^{\rm g}$  & 15:40 & no data   & no data         & no data         & no data       & [II], III$^{\rm w}$    \\
010310           & 04:00 & III       & III             & II, III         & II, III       & [II$^{\rm w}$], III           \\
010329           & 09:57 & DCIM, [III] & IV            & II, III, IV     & II, III, IV   & II?, III        \\
010402           & 10:58 & DCIM      & III?            & III, IV         & II$^{\rm FH}$, IV    & II, III$^{\rm w}$  \\
010402           & 21:32 & IV        & III, IV         & II, III, IV     & II$^{\rm FH}$, III   & II, III         \\
010409$^{\rm c}$  & 15:20 & [DCIM]    & IV              & II$^{\rm FH}$, III, IV &  II$^{\rm FH}$, III, IV  & II, III   \\
010410$^{\rm c,g}$& 05:06 & [IV]      & III, IV         & III, IV         & II, III, IV    & [II], III           \\
010412           & 09:39 & DCIM      & IV              & II$^{\rm FH}$, IV& II, III, IV   & II?, III        \\
010414$^{\rm i}$  & 17:15 & no data   &  no data        & [III]           & [II, III]      & III           \\
010415           & 13:19 & DCIM      & [IV]            & [IV]            & [II, III, IV]  & II$^{\rm FH}$, III                \\
010426           & 11:26 & DCIM      & no data         & no data         & no data        & II?, III$^{\rm w}$ \\
010719           & 09:52 & DCIM      & no data         & III$^{\rm w}$    & III$^{\rm w}$  & III$^{\rm w}$    \\
010912           & 21:05 & no data   & III             & III             & [II]           & III$^{\rm w}$                 \\
010915           & 11:04 & no data   & no?             & III             & III, IV?       & II$^{\rm FH}$, III$^{\rm w}$   \\
011019           & 00:47 & IV        & IV              & II, III, IV     & II, III        & II$^{\rm FH}$, III       \\
011019           & 16:13 & no data   & no data         & II, III, IV     & II, III, IV    & II$^{\rm FH}$, III       \\
011022           & 00:22 & III?      & III             & III             & III            & III             \\
011025$^{\rm n}$  & 14:42 & [DCIM]    & [II, IV]        & [II, IV]        & [II], III?, IV & II$^{\rm FH}$, III$^{\rm LHo}$       \\
011104$^{\rm g}$  & 16:03 & no data   & no data         & II, III, IV     & II, III, IV    & II, III       \\
011122$^{\rm g}$  & 20:18 & IV$^{\rm d}$  & bad quality  & II, III         & II, III        & [II], III           \\
011122           & 22:32 & III, IV      & III, IV      &  III, IV        &  III, IV       & [II], III          \\
011226$^{\rm g}$  & 04:32 & IV$^{\rm d}$  & IV           & III, IV         & II, III, IV    & II$^{\rm FH}$, III       \\
\hline
\end{tabular}
\end{table}

\begin{table}[ht!]
\addtocounter{table}{-1}
\caption[]{cont'd}
\label{T-Western_events2}
\tiny
\vspace{0.1cm}
\begin{tabular}{lllllll}
\hline
Event            & SXR   & \multicolumn{2}{c}{dm-$\lambda$} & \multicolumn{2}{c}{m-$\lambda$} & DH-$\lambda$  \\
yymmdd           & onset &3$-$0.8 GHz    & 0.8$-$0.3 GHz    & 300$-$100 MHz   & 100$-$30 MHz  & 30$-$0.02 MHz      \\
\hline
020220           & 05:52 & III            & III             & II, III         & II$^{\rm FH}$, III  & III           \\
020315$^{\rm c}$  & 22:09 & IV$^{\rm FS}$   & [IV]            & III, IV         & III$^{\rm w}$, [IV] & [II$^{\rm FH}$], III$^{\rm LFo}$   \\
020411           & 16:16 & DCIM           & [DCIM]          & [III]           & [III]              & [II, III]     \\
020414           & 07:28 & [DCIM, III]    & [DCIM, II], III & III             & [II], III          & II, III        \\
020415$^{\rm s}$  & 02:46 &  UNCLF         & CONT?           & III, CONT?      & III                & II, III        \\
020417           & 07:46 & DCIM, [III]    & [II], III?, IV  & [II], III, IV   & II, III, IV         & II?, III, IV   \\
020421$^{\rm g}$  & 00:43 & IV             & III, IV         & II?, III, IV    & II$^{\rm FH}$, III   & [II], III             \\
020715$^{\rm c}$  & 19:59 & IV$^{\rm FH}$   & III             & III, IV         & III, IV             & [II], III              \\
020803$^{\rm i}$  & 18:59 & no data        & no data         & III             & III                 & II$^{\rm FH}$, III   \\
020814$^{\rm g,l}$& 01:47 & UNCLF          & UNCLF           & II$^{\rm FH}$, III & II$^{\rm FH}$, III  & [II], III      \\
020816           & 05:46 & [DCIM]         & DCIM, [II$^{\rm HFo}$], III  & II$^{\rm FH}$, III & II, III & II, III$^{\rm w}$   \\
020818$^{\rm i}$  & 21:12 & no data        & III             & III             & II, III, IV         & III, IV$^{\rm d}$?  \\
020819$^{\rm i}$  & 10:28 & DCIM, [III]    & III             & III             & III                 & III               \\
020820$^{\rm i}$  & 08:22 & DCIM           & III             & III             & III                 & III             \\
020822           & 01:47 & CONT           & III             & III, IV         &  II, III, IV        & III                \\
020824$^{\rm g}$  & 00:49 & III?           & III$^{\rm d}$    & II$^{\rm FH}$, III & III, [IV]         & II$^{\rm FH}$, III          \\
021109$^{\rm g}$  & 13:08 & DCIM           & [IV]            & [IV]            & II, III, [IV]       & II?, III        \\
021219$^{\rm c}$  & 21:34 & no data        & no?             & [III, IV]       & II, III, IV         & II?, III        \\
021222           & 02:14 & IV$^{\rm d}$?   & III?            & III?            & II$^{\rm FH}$, III   & III             \\
030317           & 18:50 & no data        & no data         & III             & III                 & III              \\
030318           & 11:51 & DCIM           & [II, IV]        & [II, IV]        & II, III             & [II], III              \\
030423           & 00:39 & IV$^{\rm FH}$   & III, IV         & II$^{\rm FH}$, III, IV & II, III, IV?  & III              \\
030424           & 12:45 & DCIM           & [II], III, [IV] & [II], III, [IV] & [II], III, IV       & III$^{\rm w}$     \\
030527           & 22:56 & III, IV        & IV              & II$^{\rm FH}$, III, IV & II?, III      & [II], III              \\
030528           & 00:17 & IV             & III, IV         & III, IV         & III, IV             & [II], III, [IV]   \\
030531           & 02:13 & IV             & III, IV         & II$^{\rm FH}$, III & II$^{\rm FH}$, III & II?, III       \\
030819           & 07:38 & DCIM           & [DCIM]          & II$^{\rm FH}$, III, IV  & II$^{\rm FH}$, III, IV  & III   \\
031026           & 17:21 & no data        & no data         & no data         & II?, III, CONT?     & II?, III           \\
031029$^{\rm c}$  & 20:37 & IV             & III, IV         & II$^{\rm FH}$, III, IV   & II$^{\rm FH}$, III, IV & II?, III, IV    \\
031102           & 17:03 & [DCIM]         & no data         & no data         &  no data            & II$^{\rm FH}$, III, [IV]   \\
031103$^{\rm n}$  & 01:09 & UNCLF          & IV              & [II], III, IV   &  [II], III, [IV]    & [II], III       \\
031103$^{\rm n}$  & 09:43 & DCIM           & DCIM, [II$^{\rm HFo}$], III?  & II, III, IV  &  II, III, IV   & II$^{\rm FH}$, III, [IV]   \\
031104           & 19:29 & III, IV$^{\rm d}$ & III, IV$^{\rm d}$ & II$^{\rm FH}$, III, IV  & II, III, IV & II?, III   \\
031120$^{\rm c}$  & 07:35 & DCIM?          & III$^{\rm w}$?   & III$^{\rm w}$?   & III, IV?            & III             \\
040204           & 11:12 & DCIM           & III, IV         & II$^{\rm FH}$, III, IV & III, IV       & III$^{\rm w}$    \\
040411           & 03:54 & no?            & III             & III             & III                 & [II], III              \\
040713           & 00:09 & III            & III             & II, III         & II, III             & II?, III        \\
040725           & 14:19 & DCIM           & CONT?, III?     & III?, CONT?     & [II], III, IV       & II?, III, IV    \\
040919           & 16:46 & no data        & no data         & no data         & II, III, [IV]       & [II], III       \\
041030           & 06:08 & III            & III             & II, III         & II, III             & III              \\
041030           & 11:38 & DCIM           & [DCIM], III     & II, III, IV     & II, III, IV         & III             \\
041030           & 16:18 & no data        & no data         & no data         & II?, III            & III             \\
041107           & 15:42 & [DCIM]         & no data         & no data         & III, IV             & II$^{\rm FH}$, III, [IV]    \\
041109           & 16:59 & no data        & no data         & no data         & II, III, IV         & [II], III, [IV]              \\
041110           & 01:59 & III, IV$^{\rm FS}$ & III, IV      & II$^{\rm FH}$, III, IV  & II, III      & II?, III, IV?   \\
050115$^{\rm c}$  & 22:25 & no data        & IV              & IV              & [II], III, IV       & II, III, IV  \\
050117           & 06:59 & no data        & IV              & III, IV         & III, IV             & II, III, [IV]         \\
050119$^{\rm n}$  & 08:03 & [DCIM]         & [DCIM]          & II, III, IV     & II, III, IV         & [II], III$^{\rm w}$, [IV]  \\
050120           & 06:36 & no data        & [III], IV       & III, IV         & II, III, IV         & II, III          \\
050506           & 03:05 & III            & III             & III             & III                 & III              \\
050506           & 11:11 & DCIM, III?     & [DCIM], III?    & III             & III, IV?            & III              \\
050511           & 19:22 & no data        & no data         & no data         & II? III             & II?, III, [IV]         \\
050616           & 20:01 & no data        & no data         & III             &[II], III, [IV]      & [II], III        \\
050709           & 21:47 & no data        & IV              & III, IV         & II, III, IV         & II, III           \\
050712           & 15:47 & III?, DCIM     & III?, DCIM      & III?, [IV]      & III, [IV]           & III    \\
050713           & 02:35 & IV?            & IV?             & [III], IV?      & III$^{\rm w}$        & III$^{\rm LFo}$   \\
050713           & 14:01 & DCIM           & IV?             & IV?             & III                 & [II], III              \\
050714$^{\rm l}$  & 10:16 & [DCIM], III?   & [DCIM], III?    & III, IV         & III$^{\rm w}$, [IV]  & II, III$^{\rm LFo}$          \\
050822           & 00:44 & IV$^{\rm d}$    & IV$^{\rm d}$     & II, IV$^{\rm d}$ & II, III, IV         & II, III, [IV$^{\rm w}$]          \\
050822           & 16:46 & [IV]           & III?, IV?       & III, IV         & III, IV             & II?, III          \\
050915$^{\rm n}$  & 08:30 & DCIM           & CONT            & no?             & no?                 & no?          \\
060706           & 08:13 & [DCIM]         & II?, IV          & II, IV          & II, III, IV         & II?, III          \\
061213           & 02:14 & DCIM           & IV              & II$^{\rm FH}$, III, IV   & II$^{\rm FH}$, III, IV & II?, III, [IV]  \\
061214           & 21:07 & no data        & IV              & IV              & II$^{\rm FH}$, III, IV & [II], III, IV?                 \\
\hline
\end{tabular}
\end{table}

\begin{table}[ht!]
\caption[]{Solar energetic particle events with origin at eastern or uncertain (with superscript `u') heliolongitudes: visual identification and [observatory reports] of type II, III and IV radio bursts.}
\label{T-Eastern_events}
\tiny
\vspace{0.1cm}
\begin{tabular}{lllllll}
\hline
Event            & SXR   & \multicolumn{2}{c}{dm-$\lambda$} & \multicolumn{2}{c}{m-$\lambda$} & DH-$\lambda$  \\
yymmdd           & onset &3$-$0.8 GHz    & 0.8$-$0.3 GHz    & 300$-$100 MHz   & 100$-$30 MHz  & 30$-$0.02 MHz      \\
\hline
970401           & 13:43 &  DCIM         &  [IV]            &  [IV]           &  [II], III, [IV] & II, III             \\
970924           & 02:43 &  UNCLF        &  CONT?           &  III            &  II, III       &  III                \\
980429           & 16:06 &  DCIM         &  IV              &  III, IV        &  II, III, IV   & II?, III, IV?       \\
980818$^{\rm l}$  & 22:10 &  UNCLF        &  UNCLF           &  II, III        &  II, III, [IV] & II, III$^{\rm w}$, IV?   \\
980819           & 21:35 &  UNCLF        &  UNCLF           &  II, III        &  II, III, [IV] & II?, III$^{\rm w}$, IV?   \\
980824$^{\rm c,g}$& 21:50 &  IV?          &  III?, IV?       &  II, III, [IV]  &  II$^{\rm FH}$, III, [IV] & II$^{\rm FH}$, III             \\
980920           & 02:33 &  no?          &  no?             &  III            &  II, III, IV   & [II], III           \\
980923$^{\rm c,g}$& 06:40 &  DCIM         & IV               &  II, IV         &  II, IV        & II, III             \\
990503$^{\rm g}$  & 05:36 &  DCIM         & IV               &  IV             &  II, III, IV   & II$^{\rm FH}$, III             \\
990629$^{\rm c}$  & 05:01 &  no?          & [DCIM, III]      & [DCIM], III     &  II, III       & II, III             \\
991117           & 09:47 &  DCIM         & no?              &  no?            &  III           & II$^{\rm FH}$, III             \\
000118           & 17:07 &  no data      & no data          &  no data        &  [II, III, IV] & [II], III            \\
000217$^{\rm c}$  & 20:17 &  no data      & no?              &  II             &  II, III       & II$^{\rm FH}$, III           \\
000510           & 19:26 & no data       & no?              &  UNCLF          &  II, III, IV   & III                 \\
000606$^{\rm g}$  & 14:58 &  DCIM         & [II$^{\rm HFo}$, III]  & [II, III] &  II, III, IV   & II$^{\rm FH}$, III             \\
000710           & 21:05 &  no data      & CONT             &  IV             &  II, III, IV   & II, III, [IV]     \\
001029           & 01:28 &  no data      & IV               &  IV             &  II, III, IV   & II, III      \\
001125$^{\rm s}$  & 00:59 &  no data      & IV               &  IV             &  II, III, IV   & II?, III            \\
010120           & 18:33 &  no data      & no data          &  II, III        &  II, III       & II$^{\rm FH}$, III             \\
010325           & 16:25 &  [III, IV]    & no data          &  no data        &  IV?           & III                 \\
010615           & 10:01 &  DCIM         & [DCIM]           & [DCIM, II], III &  III, IV       & [II$^{\rm FH,w}$], III                 \\
010917$^{\rm c}$  & 08:18 &  DCIM         & no data          &  II, III        &  II, III       & II, III             \\
010924$^{\rm g}$  & 09:32 &  DCIM         & no data          &  III, IV        &  II$^{\rm FH}$, III, IV & II, III     \\
011009           & 10:46 &  DCIM         & no data          &  IV?            &  II, III, IV   & II, III             \\
011022           & 14:27 &  [DCIM]       & [IV]             &  [III]          &  II, III, IV   & II, III             \\
011117$^{\rm g}$  & 04:49 &  no data      & IV               &  III, IV        &  II, III, IV   & II?, III            \\
011128           & 16:26 &  no data      & no data          &  no data        &  II, III, IV   & II?, III             \\
020520           & 15:21 &  DCIM, III    & [II], III, [IV]  &  [II, III, IV]  & [II, III, IV]  & III              \\
020816           & 11:32 &  DCIM         & [II$^{\rm HFo}$], III, IV & [II], III, IV &  II?, III, IV  & II, III, IV      \\
030421$^{\rm c}$  & 12:54 &  III?, DCIM   &  III?, [IV]      &  II, III, IV    &  II, III, IV   & III, IV?    \\
030425           & 05:23 &  [III]        &  [III]           &  II, III        &  II, III       & II, III        \\
030615           & 23:25 &  no data      &  III, IV?        &  II, III, IV    &  II, III, IV   & II, III          \\
030717           & 08:17 &  DCIM         &  III             &  II, III        &  II, III       & II?, III          \\
031026           & 05:57 &  DCIM         &  IV              &  [II], III, IV  &  II, III, IV   & II, III, IV?     \\
031028$^{\rm c}$  & 09:51 &  DCIM, [III]  &  [II], III, IV   &  [II] III, IV   &  II, III, IV   & II, III, IV      \\
031118           & 07:23 &  DCIM         &  [II$^{\rm HFo}$], III, IV  &  II, III, [IV]  &  II, III, IV   & II$^{\rm w}$, III, [IV]     \\
040107           & 10:14 &  DCIM         &  [IV]            &  II, III, [IV]  &  II$^{\rm FH}$, III & II$^{\rm FH}$, III        \\
040912           & 00:04 &  no data      &  IV              &  III, IV        &  II$^{\rm FH}$, III, IV & II?, III          \\
041104           & 22:53 &  no data      &  III, IV         &  III, IV        &  [II], III, IV & II?, III, IV?     \\
041202$^{\rm c}$  & 23:44 &  IV           &  IV              &  IV             &  II, III, IV   & II, III          \\
050114$^{\rm c}$  & 10:08 &  DCIM         &  [DCIM], III     &  III, IV        &  [II], III, IV & III              \\
050115$^{\rm c}$  & 05:54 &  IV?          &  IV              &  IV             &  II, III, IV   & II, III, IV      \\
050513           & 16:13 &  DCIM         &  III?, [IV]      &  [II, IV]       &  [II, III, IV] & II?, III, IV?    \\
050603$^{\rm l}$  & 11:51 &  DCIM         &  IV              &  [III]          &  [II, III]     & II, III          \\
050907$^{\rm l}$  & 17:17 &  no data      &  no data         &  no data        &  [II, IV]      & II, III          \\
050913           & 19:19 &  no data      &  no data         &  no data        &  [II], III, IV & II, III, IV      \\
061106$^{\rm l}$  & 17:43 &  no data      &  no data         &  [II], III      &  [II], III     & II, III          \\
061205           & 10:18 &  DCIM         &  [DCIM], III     & [DCIM], II, III &  II, III       & II, III          \\
061206           & 18:29 &  no data      &  no data         &  [II], III, [IV]& [II], III, [IV]& II, III          \\
\hline
980909$^{\rm u}$  & 04:52 &  no?          & no?              &  II, III        & II, III        & III            \\
990216$^{\rm u}$  & 02:49 &  CONT         & CONT             &  II?, IV?       & [II], III      & III             \\
\hline
\end{tabular}
\end{table}

\section{Results}

\begin{figure}[t!]
   \centerline{\hspace*{-0.05\textwidth}
               \includegraphics[width=0.5\textwidth,clip=]{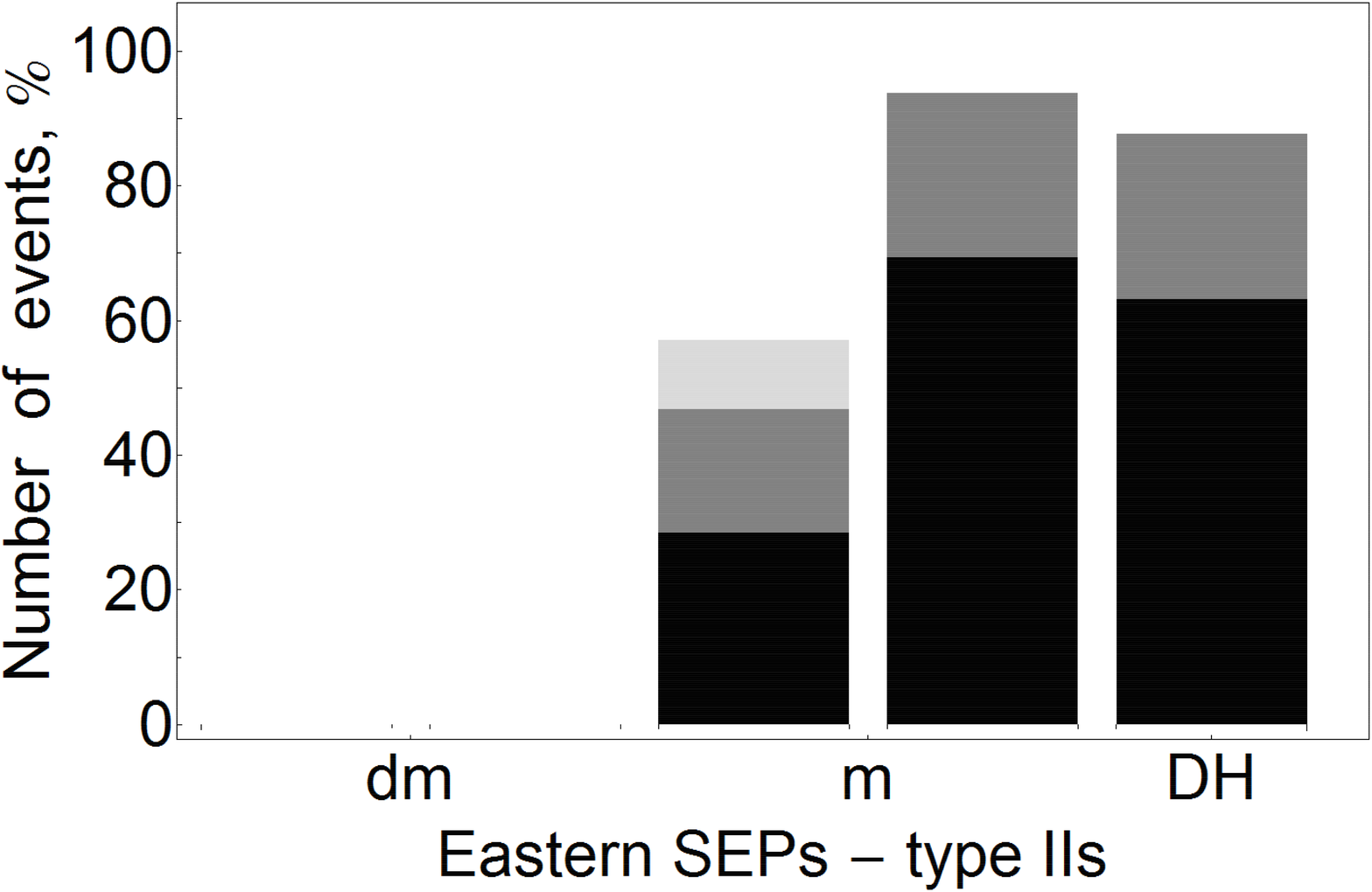}
               \hspace*{-0.01\textwidth}
               \includegraphics[width=0.5\textwidth,clip=]{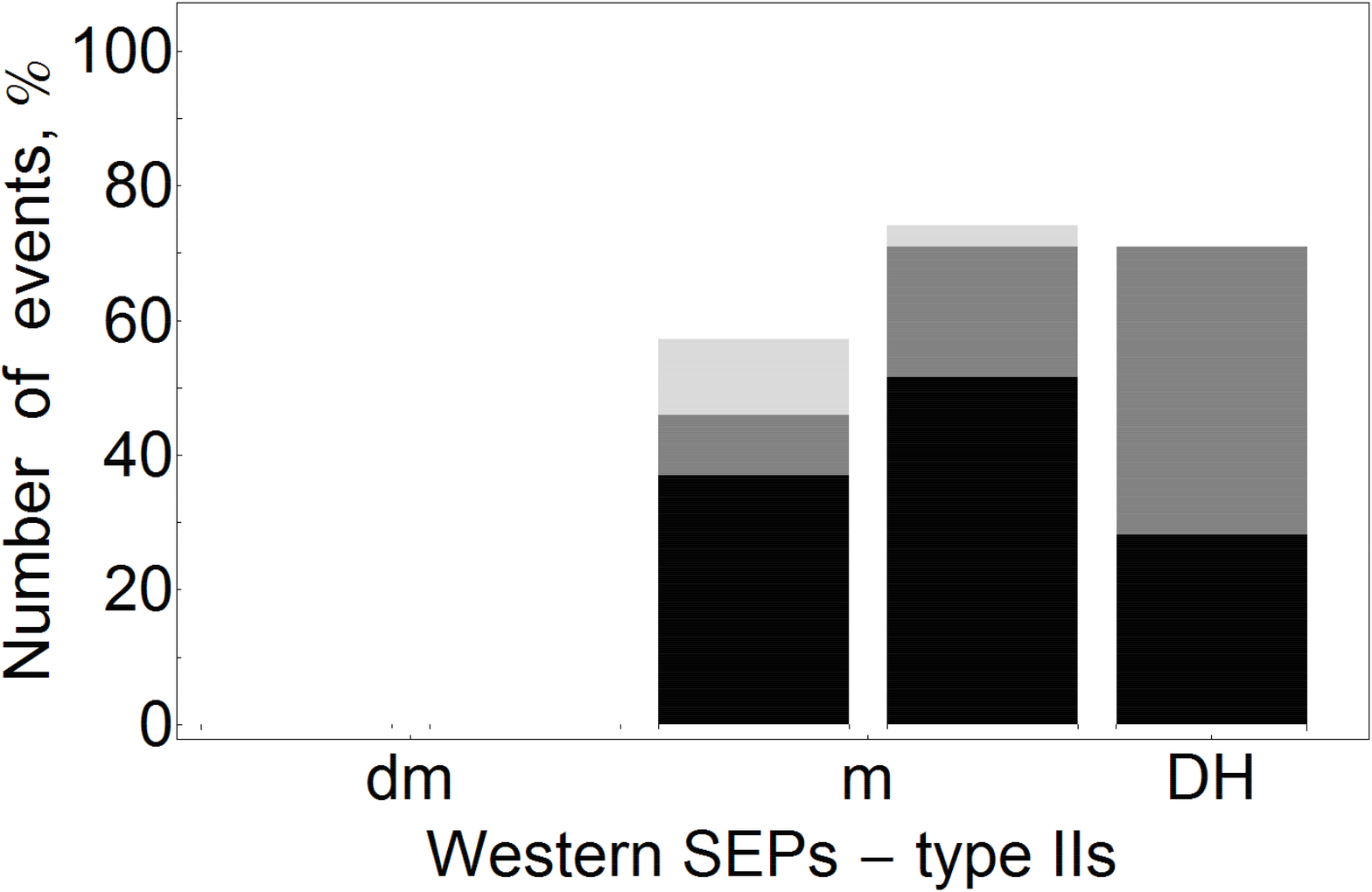}
               }
               \vspace*{-0.0\textwidth}
   \centerline{\hspace*{-0.05\textwidth}
               \includegraphics[width=0.5\textwidth,clip=]{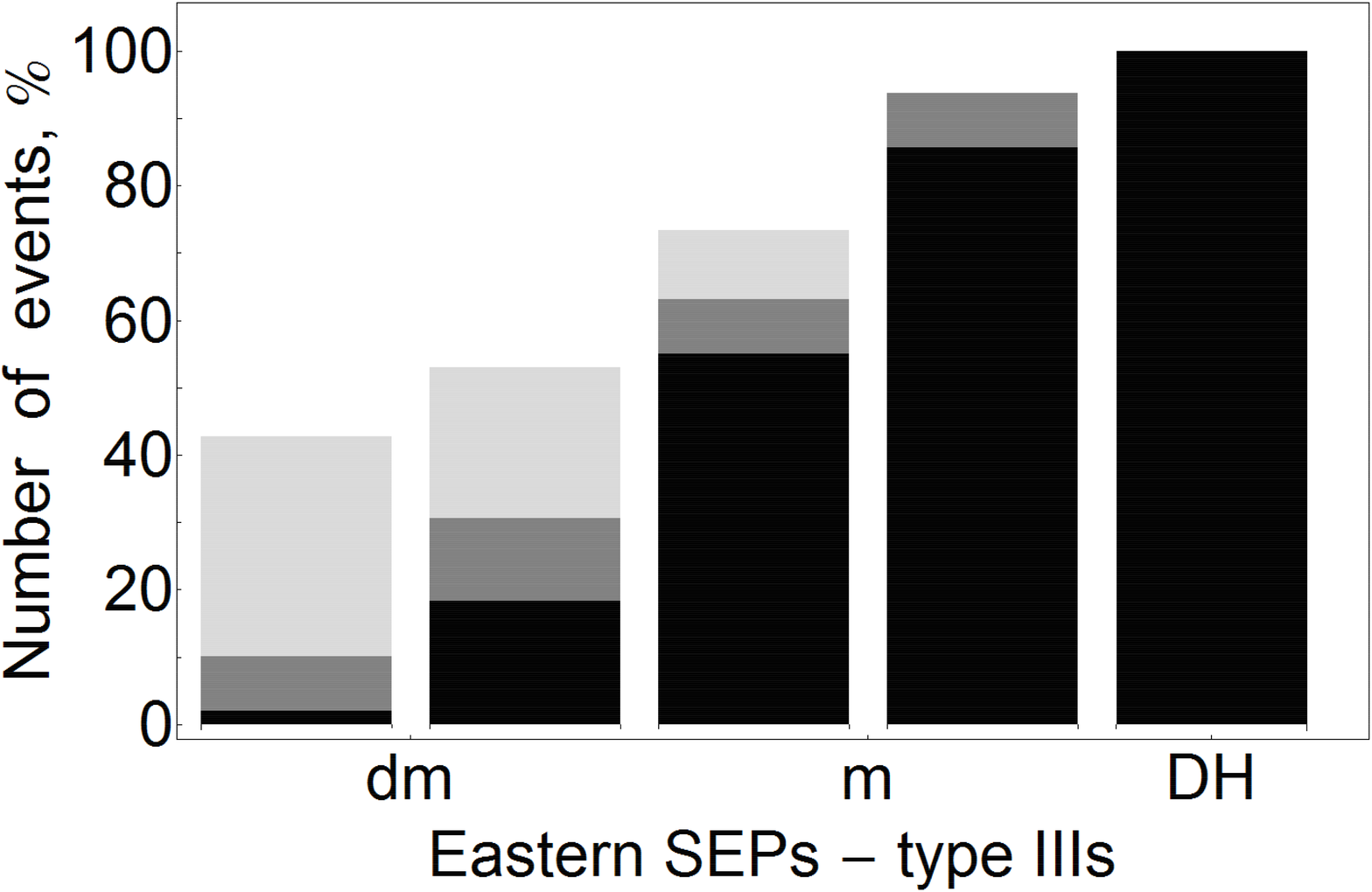}
               \hspace*{-0.01\textwidth}
               \includegraphics[width=0.5\textwidth,,clip=]{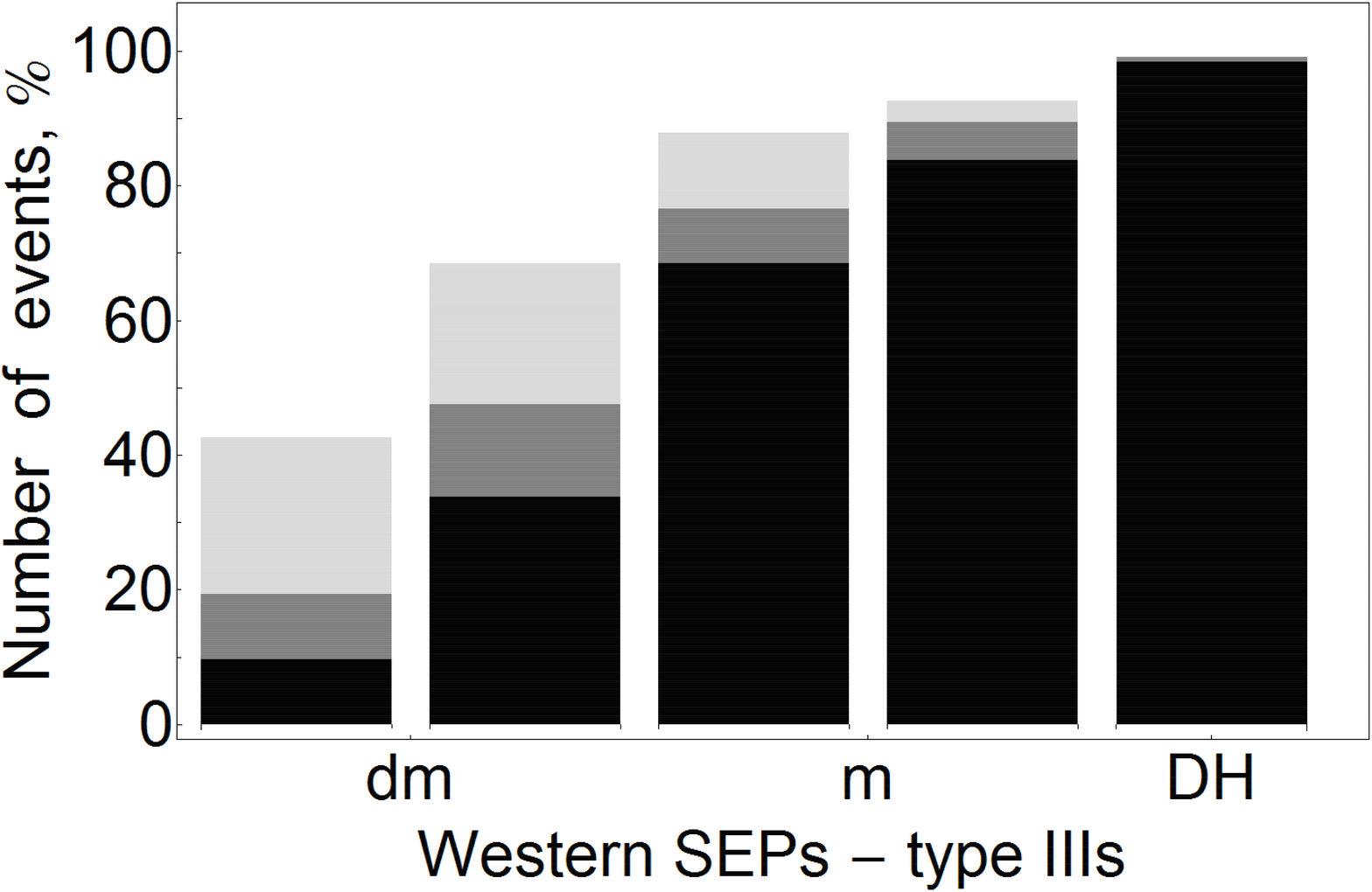}
               }
               \vspace*{-0.0\textwidth}
   \centerline{\hspace*{-0.05\textwidth}
               \includegraphics[width=0.5\textwidth,clip=]{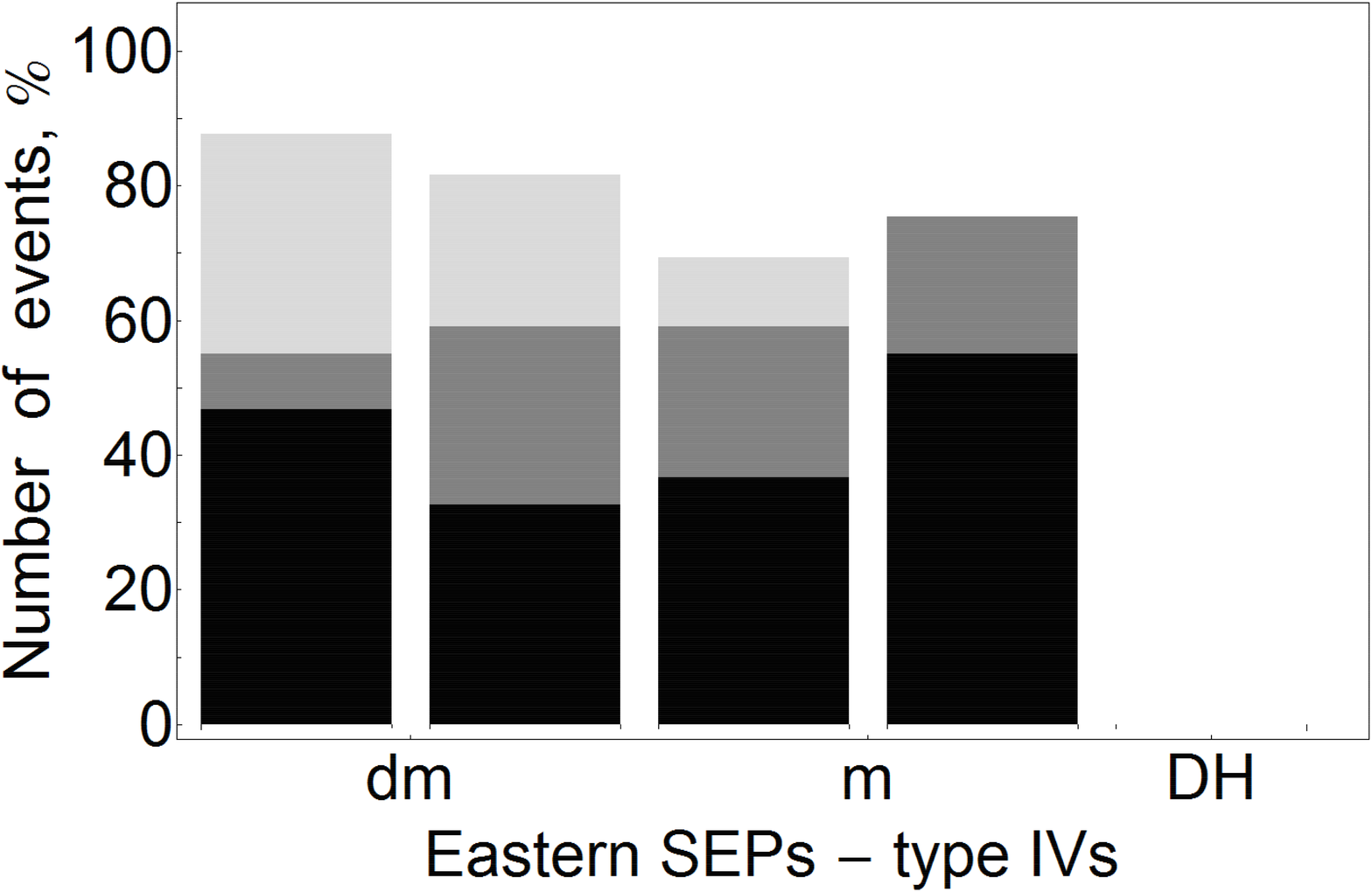}
               \hspace*{-0.01\textwidth}
               \includegraphics[width=0.5\textwidth,clip=]{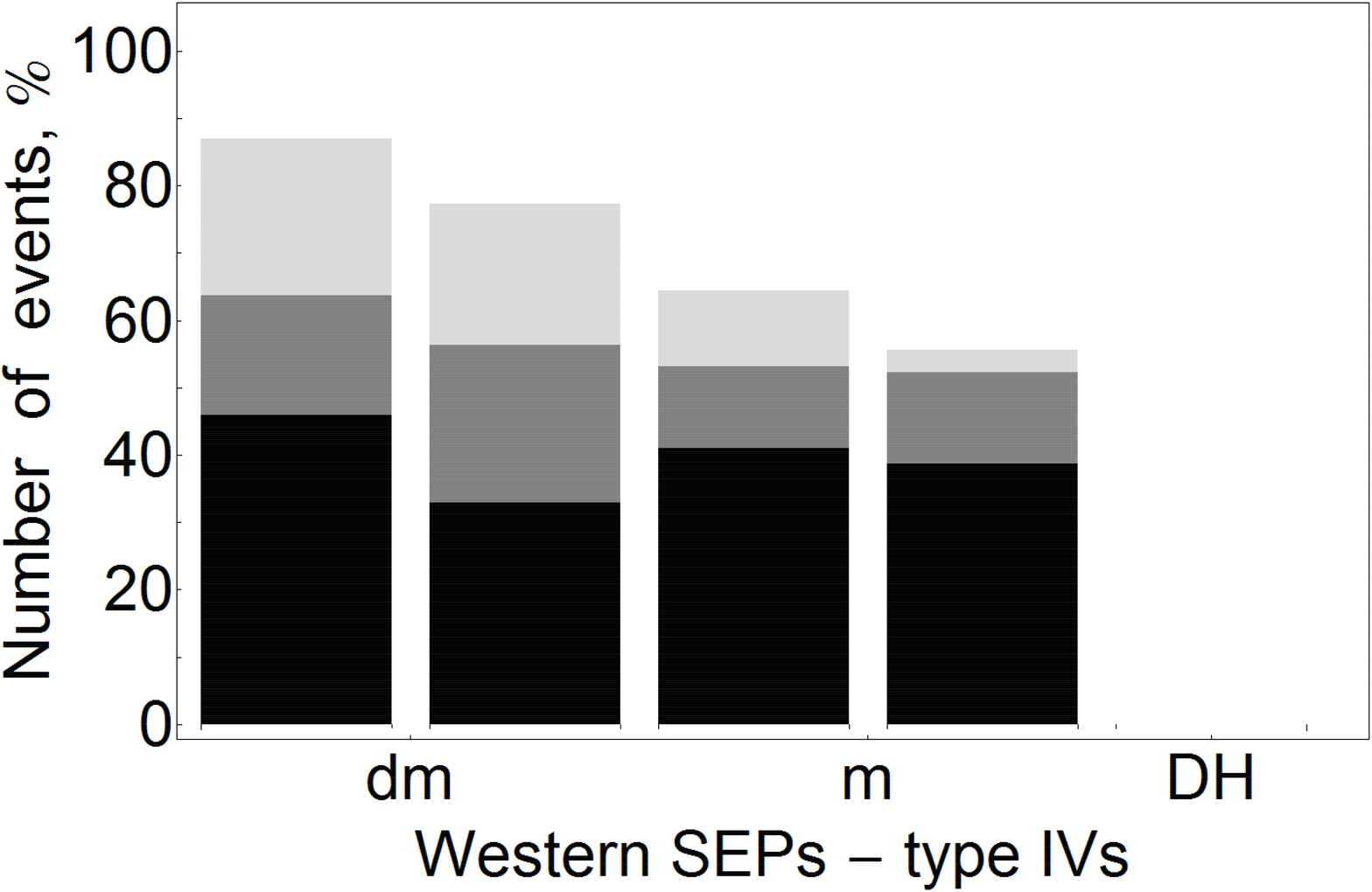}
               }
\caption{Histograms of the association rates of eastern (lest) and western (right) SEP events and the corresponding radio burst types for dm (3$-$0.8 and 0.8$-$0.3 GHz), m (300$-$100 and 100$-$30 MHz) and DH (30$-$0.02 MHz) range. For the color-code see text.}
   \label{F-histo}
   \end{figure}

The results are given as normalized number of the SEP events vs. radio frequency for each wavelength range (dm, m and DH), see the histograms on Figure~\ref{F-histo}, for eastern (on the left) and western (right) SEP events. The numerical value of the association rates, graphically presented on the histograms, is given by the height of each color bar. The SEP events associated with specific burst types as identified on the radio spectral plots are given with black color. With dark gray is shown the association in cases where the burst type was only given in the observatory reports (no spectra found at present) or where its identification is questionable. Whenever we give a value for an association rate, we will always sum up these two sections. Finally, with light gray color we denote the number of SEP events for which no radio information could be found (neither plots nor observatory reports). The majority of the missing radio plots is in the dm-range due to poor data coverage. Note that the highest discrepancy between the results given by us and by observatory reports is for the type II burst identification in the DH-range. This is mostly due to the weak and intermittent appearance of the IP type II bursts which makes an identification on quick-look plots difficult. In addition, the subjectivity of the observer plays a prominent role here, whereas the DH-type III identification, for example, is straightforward.

On the histograms, the number of events in each column is normalized to the total number of events in each group (eastern and western, correspondingly) and is also given explicitly in Table~\ref{T-Rates_Sum}. While representing the association rate in the dm and m-range, we chose the greater association rate among their two subbands. For the total number of SEP events, given with `All' in Table~\ref{T-Rates_Sum}, we sum up the eastern, western and the uncertain SEP events. Since the dm-type II burst and the DH-type IV bursts are intrinsically rare phenomena, they will be excluded from the further analysis.

The highest association rate of SEPs is with the m and DH-type III bursts, usually $\gtrsim 90$\% (see Table~\ref{T-Rates_Sum} for details). A much lower association rate between SEPs and type IIIs is found in the dm-range (from about 30 to 50\%). No dependence on the eastern vs. western heliolongitudes is seen for type III and type IV bursts, with the association rate of the type IV bursts being in the range from 50\% up to 75\%. In contrast, the SEP-association rate with the 30$-$100 MHz metric (94\%) and the DH type II bursts (88\%) is slightly higher for eastern SEP events than for western ones (71 \%).

\begin{sidewaystable}[ht!]
\caption[]{Association rates of SEP events and solar radio bursts.}
\label{T-Rates_Sum}
\tiny
\vspace{0.5cm}
\begin{tabular}{lccccccclcc}
\hline
\\
SEP     &  \multicolumn{7}{c}{Association rates, \%}                     & Reference     & Time     & Number \\
\\
category& m-II  & DH-II   & dm-III &  m-III &  DH-III & dm-IV   & m-IV   &               & coverage & of events \\
\\
\hline
& \\
& \multicolumn{7}{c}{{\bf SEP events $-$ radio bursts}} &\\
& \\
Eastern &  94  & 88 & 31 & 94 & 100 & 59 & 75                 & this work & 1997$-$2006 & 49   \\
        &    &  &  &  &  &  &      \\
Western &  71  & 71 & 48 & 89 & 99 & 64 &  53                 &  this work & 1997$-$2006 & 124 \\
        &  &  &  &  &  &  &      \\
        &      & 90$^{\rm sr}$ &  &  &  &  &  & \cite{2007ApJ...658.1349C}  & 1997$-$2003 & 140 \\
        &    &  &  &  &  &  &      \\
All     &  78  & 75 & 42 & 91 & 98 & 61 & 58                  & this work & 1997$-$2006 & 175 \\
       &  &  &  &  &  &  &      \\
        &  82 (88$^{\rm s}$/80$^{\rm w}$)  & 63 (96$^{\rm s}$/50$^{\rm w}$) &  &  &  &  &  & \cite{2004ApJ...605..902C} & 1996$-$2001 & 88 (24$^{\rm s}$/64$^{\rm w}$) \\
       &  &  &  &  &  &  &      \\
        &  90$^{\rm f,m}$     & 100$^{\rm m}$     &  &  &  &  &  & \cite{2003GeoRL..30lSEP1G} & 1997$-$2001 & 48   \\
       &  &  &  &  &  &  &      \\
        &                     &                   & \multicolumn{3}{c}{$100^{\rm l}$} &  &  & \cite{2002JGRA..107.1315C} & 1997$-$2001 & 123     \\
       &  &  &  &  &  &  &      \\
       &  &  &  &  & 91$^{\rm l,m}$ &  &  & \cite{2003GeoRL..30lSEP6M} & 1997$-$2001 &  47   \\
       &  &  &  &  &  &  &      \\
        &                     &                   &  &  &  &  &  88 &  \cite{1982ApJ...261..710K} & 1973$-$1980 & 52   \\
        &  &  &  &  &  &  &      \\
& \multicolumn{7}{c}{{\bf Radio bursts $-$ SEP events}} &\\
       &  &  &  &  &  &  &      \\
Eastern &  \multicolumn{2}{c}{8$^{\rm yN}$/52$^{\rm yY}$} &  &  &  &  &  & \cite{2004ApJ...605..902C} & 1996$-$2001 & 50/23    \\
        &    &  &  &  &  &  &      \\
Western &  \multicolumn{2}{c}{25$^{\rm yN}$/90$^{\rm yY}$}&  &  &  &  &  & \cite{2004ApJ...605..902C} & 1996$-$2001 & 69/29   \\
       &  &  &  &  &  &  &      \\
All    & \multicolumn{2}{c}{61$^{\rm yY}$/52$^{\rm nY}$/30$^{\rm yN}$} &  &  &  &  &  & \cite{2008AnGeo..26.3033G} & 1996$-$2005 & 165/69/26     \\
       &  &  &  &  &  &  &      \\
\hline
\end{tabular}
\footnotetext{s (w): strong (weak) SEP event with peak intensity of $\geq (<)\, 0.1$ cm$^{-2}$s$^{-1}$MeV$^{-1}$sr$^{-1}$; sr: stronger SEP peak flux, $> 10$ MeV protons of $3\sim 10^4$ cm$^{-2}$s$^{-1}$sr$^{-1}$; f: frontsided source location; l: long-lasting type III emission; m: major SEP event, $> 10$ MeV protons of $\geq 10$ cm$^{-2}$s$^{-1}$MeV$^{-1}$sr$^{-1}$; yN: m but no DH-IIs; yY: m and DH-IIs; nY: no m but DH-IIs.}
\end{sidewaystable}

\section{Discussion}

We present the association rates of the SEP events (protons) and their accompanying radio emission (from electrons) in the corona (dm and m wavelength) and IP space (DH-range). Since there are no signatures of protons interacting with the solar atmosphere (with the exception of gamma-ray emission), we use electron signatures as a diagnostic for particle acceleration from the corona up to 1~AU.

\cite{2002JGRA..107.1315C} were the first to identify long-lasting groups of DH type III bursts as a typical radio counterpart of large SEP events. They found the groups to be of significantly longer duration than type~III bursts associated with impulsive flares, see also \cite{1987SoPh..111..397M}. In the analysis performed here, we did not take into account the burst duration of DH-type III, nor explicitly separate the SEP events into gradual or impulsive. Still we find that SEP events have the highest association rate with type III radio bursts. This implies that the electrons accelerated in the corona (within one solar radius) have a ready access to the IP space, irrespective of whether the SEP event is impulsive or gradual.

Shock signatures (type II bursts) were also considered in correlation studies with SEP events. We found a lower association rate (up to 75\%) of the DH-type II bursts with SEP events, compared to the m- and DH-type III association rates. The number increases when only SEP events with strong intensities are considered, in agreement with \cite{2002ApJ...572L.103G} and \cite{2004ApJ...605..902C}.

The association of SEP events with types III and IV is comparable in the eastern and western groups (see Figure~\ref{F-histo}). But for the type II bursts there is a slight trend for a higher association rate in the eastern hemisphere. This result could be understood it terms of different sources contributing to SEP events. Shocks could be the dominant accelerator when the parent activity is poorly connected to Earth (as in the eastern solar hemisphere). In the western hemisphere (where more than twice as many events were detected) both flare and shock acceleration could contribute. When no shock signatures accompany the eastern SEP events, their propagation through the IP space and detection at Earth could be facilitated by a large-scale magnetic structure, e.g., interplanetary coronal mass ejections (ICMEs). \cite{1991JGR....96.7853R} estimated that for about 15\% of the eastern events this is likely the case. Recently \cite{2013SoPh..282..579M} showed similar occurrence rate of ICMEs for the western SEP events (20\%).

The association with the type IIIs seems to be increasing with the decreasing of the radio frequency (from dm to DH-range). At first, the type III identification in the dm-range might be masked due to overlying decimeter continuum often present in the dynamic radio spectra (observational bias). Hence, the obtained association rates of dm-IIIs are to be considered as lower limits only. In addition, high frequencies and dense plasma impede the production and growth of Langmuir waves. Also, the electron beam must travel an `instability distance' from the acceleration site before Langmuir waves are generated \citep{2011A&A...529A..66R}. This can cause a lack of high frequency type III emission in even the most energetic of electron beams. Moreover, collisions of both waves and electrons can suppress Langmuir wave growth at high frequencies, e.g., \citet{1982ApJ...263..423K,2012SoPh..tmp..109R}.

Without the intention to give a complete overview on the results from previous work, we selected few statistical studies that are (partially) co\-vering solar cycle 23. The different association rates are given in Table~\ref{T-Rates_Sum} and are mostly consistent, although in many earlier studies only large SEP events were considered.

\section*{Acknowledgements}
R.M. acknowledges a post-doctoral fellowship from Paris Observatory. A.N. was partly supported by the European Union (European Social Fund ESF) and Greek national funds through the Operational Program "Education and Lifelong Learning" of the National Strategic Reference Framework (NSRF) - Research Funding Program: Thales. Investing in knowledge society through the European Social Fund. H.R. asknowledges the suport of the Scottish Universities Physics Alliance. We acknowledge the open data policy for the radio data used in this study and partial funding from the European Union Seventh Framework Programme (FP7/2007-2013) under grant agreement No. 262773 (SEPServer) and HESPE network (FP7-SPACE-2010-263086).

\bibliographystyle{ceab}
\bibliography{mitevasample}

\end{document}